\documentclass[aps,prb,twocolumn,showpacs,groupedaddress]{revtex4}
\usepackage{graphicx}
\usepackage{natbib}
\usepackage[utf8]{inputenc}
\usepackage{epstopdf}
\setcitestyle{super}
\usepackage{color}
\usepackage[normalem]{ulem}

\begin{document}
	
	\title{Pressure-temperature phase diagram of CaK(Fe$_{1-x}$Mn$_x$)$_4$As$_4$ for $x$=0.024}
	\author{Li Xiang\footnote{Current affiliation: National High Magnetic Field Laboratory, Florida State University, Tallahassee, Florida 32310, USA}\footnote{li.xiang@magnet.fsu.edu}}
	\address{Ames Laboratory, US Department of Energy, Iowa State University, Ames,
		Iowa 50011, USA}
	\address{Department of Physics and Astronomy, Iowa State University, Ames, Iowa 50011,
		USA}
	\author{Mingyu Xu}
	\address{Ames Laboratory, US Department of Energy, Iowa State University, Ames,
		Iowa 50011, USA}
	\address{Department of Physics and Astronomy, Iowa State University, Ames, Iowa 50011,
		USA}
	\author{Sergey L. Bud'ko}
	\address{Ames Laboratory, US Department of Energy, Iowa State University, Ames,
		Iowa 50011, USA}
	\address{Department of Physics and Astronomy, Iowa State University, Ames, Iowa 50011,
		USA}
	\author{Paul C. Canfield\footnote{canfield@ameslab.gov}}
	\address{Ames Laboratory, US Department of Energy, Iowa State University, Ames,
		Iowa 50011, USA}
	\address{Department of Physics and Astronomy, Iowa State University, Ames, Iowa 50011,
		USA}
	\date{\today}

	\begin{abstract}
		Resistance measurements on single crystals of CaK(Fe$_{1-x}$Mn$_x$)$_4$As$_4$ ($x$ = 0.024) were performed under hydrostatic pressure up to 5.15 GPa. The pressure dependence of the magnetic and superconducting transition temperatures and that of the superconducting upper critical field are reported. Our results show that upon increasing pressure, the magnetic transition temperature $T_{N}$ is suppressed, whereas the superconducting transition temperature $T_{c}$ first increases and then decreases, exhibiting a maximum at a pressure $p_c$ corresponding to the intersection of the $T_{N}$($p$) and $T_{c}$($p$) lines. In addition, a minimum in the normalized slope of the superconducting upper critical field as well as a change in the pressure dependence of the inferred superconducting coherence length are observed at $p_c$, suggesting a difference in the Fermi surface of the paramagnetic and antiferromagnetic states. Finally, CaK(Fe$_{1-x}$Mn$_x$)$_4$As$_4$ ($x$ = 0.024) likely goes through a half-collapsed-tetragonal phase transition at $\sim$ 4.3 GPa, further demonstrating that the half-collapsed-tetragonal transition pressure in the CaKFe$_4$As$_4$ system is relatively insensitive to transition metal substitution.
	\end{abstract}
	
\maketitle
	
	\section{Introduction}
	Since their discovery, Fe-based superconductors (Fe-SC) have attract great interest for their unconventional superconductivity as well its interplay with magnetic, structural degrees of freedom\cite{Kamihara2008,Takahashi2008Nat,Rotter2008PRL,Ren2008,Paglione2010}. The intensively investigated BaFe$_2$As$_2$ family shaped the canonical picture of the phase diagram of Fe-SC\cite{Canfield2010,Paglione2010}. At ambient pressure, the parent compound is tetragonal and paramagnetic at high temperature. Upon cooling, it goes through a magnetic/structural phase transition to a stripe-type antiferromagnetic, orthorhombic state\cite{Canfield2010,Johnston2010,Dai2015}. Upon chemical substitution or physical pressurization, the magnetic/structural transitions are separated and suppressed to lower temperatures and at certain level superconductivity emerges, often in a dome-like region with the magnetic/structural ordering line intersecting the $T_ c$-dome near its maximal value\cite{Paglione2010,Torikachvili2008PRB,Kimber2009,Avci2012,Wu2013a}.
	
	Recently, a new class of Fe-SC, the $AeA$Fe$_4$As$_4$ ($Ae$ = Alkaline Earth, $A$ = K, Rb, Cs), was discovered\cite{Iyo2016}. In the $AeA$Fe$_4$As$_4$ family, $Ae$ and $A$ form alternating planes along the $c$ axis, which distinguishes $AeA$Fe$_4$As$_4$ from the chemically substituted ($Ae_{0.5}A_{0.5}$)Fe$_2$As$_2$, where $Ae$ and $A$ randomly occupy the same site (e.g. Ba$_ 0.5$K$_ 0.5$Fe$_2$As$_2$)\cite{Iyo2016}. Although these $AeA$Fe$_4$As$_4$ compounds were discovered in the polycrystalline form, single crystals of CaKFe$_4$As$_4$ were synthesized and investigated\cite{Meier2016PRB,Meier2017PRM}. Pure, or undoped, CaKFe$_4$As$_4$ is superconducting below $\sim$ 35 K without any other structural or magnetic phase transitions. Upon pressure tuning up to $\sim$ 6 GPa, the superconducting transition temperature $T_ c$ is weakly suppressed and a half-collapsed-tetragonal (hcT) transition that destroys bulk superconductivity is identified at $\sim$ 4 GPa\cite{Kaluarachchi2017PRB}.
	
	On one side, electron doping of CaKFe$_4$As$_4$ via Ni- or Co-substitution suppresses superconductivity and induces a new-type of antiferromagnetic order, the hedgehog-spin-vortex magnetic order that has no accompanying structural transition\cite{Meier2018}. Pressure tuning of the Ni-substituted CaKFe$_4$As$_4$ up to $\sim$ 5 GPa leads to a suppression of the $T_ N$ associated with the hedgehog-spin-vortex antiferromagnetic transition and a small local maximum in $T_ c$ located where $T_ c$($p$) and $T_ N$($p$) lines intersect and a hcT transition at $\sim$ 4 GPa\cite{Xiang2018PRB}. On the other side, a very recent study reports that nominal hole doping into CaKFe$_4$As$_4$ via Mn substitution also reveals a magnetic phase region as well as suppressing the superconducting transition temperature\cite{Xu2022PRB}. In the BaFe$_2$As$_2$ family, the effects of the transition metal substitution have been studied intensively and it is suggested that compared with Co- or Ni-substitution, Mn behaves more local-moment-like. For example, solely Mn substitution could not induce superconductivity in BaFe$_2$As$_2$ as compared to Co or Ni substitution\cite{Thaler2011,Canfield2010}. Mn substitution suppresses superconducting transition temperature more rapidly than Co or Ni\cite{Li2012,Xu2022PRB}. 
	
	The observation that Mn appears to behave more local-moment-like than Ni or Co\cite{Xu2022PRB} has its consequences for coexistence of superconductivity and magnetism in these materials. Whereas in Ni-substituted CaKFe$_4$As$_4$, the coexistence of SC and magnetism is fairly consistent with a simple model\cite{Machida1981} in which SC and itinerant magnetism compete for the same electronic density of states\cite{Budko2018,Kreyssig2018}. In Mn-substituted CaKFe$_4$As$_4$, Mn appears to have local-moment-like behavior, so that simple model of Machida does not work any more, and, in part, the observed $T_c$ suppression with Mn\cite{Xu2022PRB} is due to Abrikosov-Gor'kov mechanism\cite{Abrikosov1961}. So it appears that Mn-substituted CaKFe$_4$As$_4$ has enough significant differences with Ni (/Co)  substituted ones to investigate the tunability of the magnetic phase and its interplay with superconductivity.
	
	In this work, we present a pressure study of CaK(Fe$_{1-x}$Mn$_x$)$_4$As$_4$ ($x$ = 0.024) up to $\sim$ 5 GPa. From resistance measurements, the superconducting transition temperature, $T_ c$ and the magnetic transition temperature, $T_ N$, were identified. A pressure-temperature ($p-T$) phase diagram is constructed accordingly. Upon increasing pressure, $T_ c$ first increases and then decreases with a maximum at $\sim$ 2 GPa, whereas $T_ N$ is monotonically suppressed intersecting the $T_ c$($p$) line at $p_c \sim$ 2 GPa. Pressure-dependent resistance analysis also suggest a hcT transition at $\sim$ 4.3 GPa. Furthermore, superconducting upper critical field analysis suggests a Fermi-surface reconstruction as well as a change in the pressure dependence of the superconducting coherence length at $p_c$.
	
	\section{Experimental Details}
	Single crystals of CaK(Fe$_{1-x}$Mn$_x$)$_4$As$_4$ (x = 0.024) with sharp superconducting transitions at ambient pressure (see Figs. \ref{fig1}(b) and \ref{fig2}(b)) were grown using high-temperature solution growth\cite{Meier2016PRB,Meier2017PRM}. The substitution level x was determined by performing Energy dispersive x-ray spectroscopy (EDS)\cite{Xu2022PRB}. The in-plane $ab$ resistance was measured using the standard four-probe configuration. The 25 $\mu$m Pt wires were spot weldered to the samples. Two samples, 1 and 2, were measured in a piston-cylinder cell (PCC)\cite{Budko1984} and a modified Bridgman anvil cell (MBAC)\cite{Colombier2007}, respectively. Pressure values for both cells, at low temperature, were inferred from the $T_c(p)$ of lead\cite{Bireckoven1988,Xiang2020}. For the PCC, a 4:6 mixture of light mineral oil:n-pentane was used as the pressure medium, which solidifies, at room temperature, in the range of 3–4 GPa. For the MBAC, a 1:1 mixture of isopentane:n-pentane was used as the pressure medium, which solidifies, at room temperature, in the range of 6–7 GPa. Both of the solidification pressures are well above the maximum pressures achieved in the pressure cells, which suggests good hydrostatic conditions\cite{Kim2011,Torikachvili2015}. The ac resistance measurements were performed in a Quantum Design physical property measurement system (PPMS) using I = 1 mA; f = 17 Hz excitation, on cooling with a rate of 0.25 K/min, and the magnetic field was applied along the crystallographic $c$-axis of the CaK(Fe$_{0.976}$Mn$_{0.024}$)$_4$As$_4$ samples.
	
	\section{Results and Discussion}
\begin{figure}
		\begin{center}
			\includegraphics[width=\columnwidth]{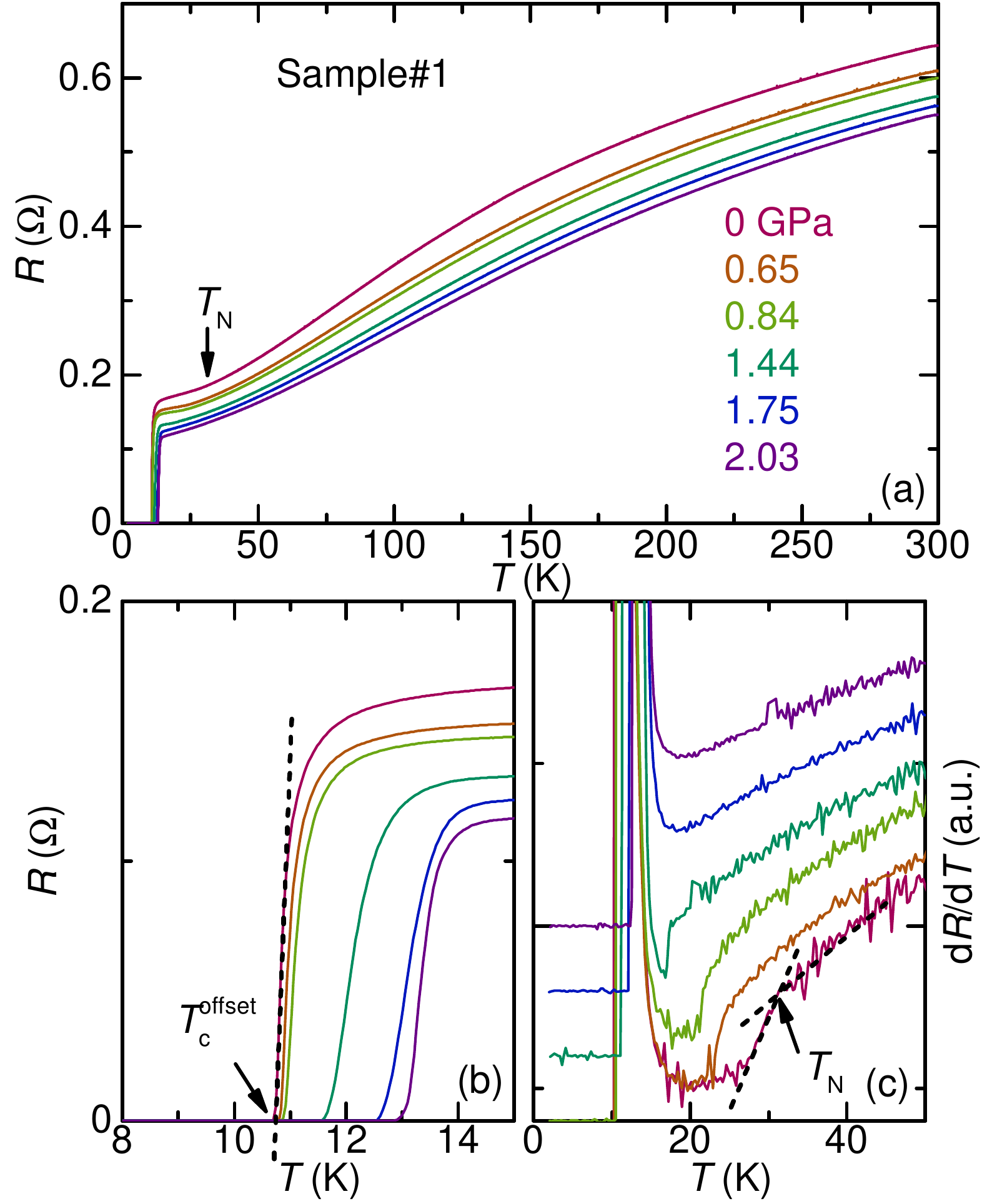} 
			\caption{(a) Evolution of the in-plane resistance with hydrostatic pressure up to 2.03 GPa measured in a piston-cylinder cell (PCC) for CaK(Fe$_{0.976}$Mn$_{0.024}$)$_4$As$_4$, sample 1. (b) Enlarged view of the low-temperature resistance showing the superconducting transition. Criterion for $T_c^{offset}$ is indicated by dashed line and arrow in the figure. (c) Temperature derivative of the resistance, d$R$/d$T$, showing the evolution of the magnetic transition $T_{N}$. Criterion for $T_{N}$ is indicated by dashed lines and arrow in the figure.}
			\label{fig1}
		\end{center}
	\end{figure}

\begin{figure}
		\begin{center}
			\includegraphics[width=\columnwidth]{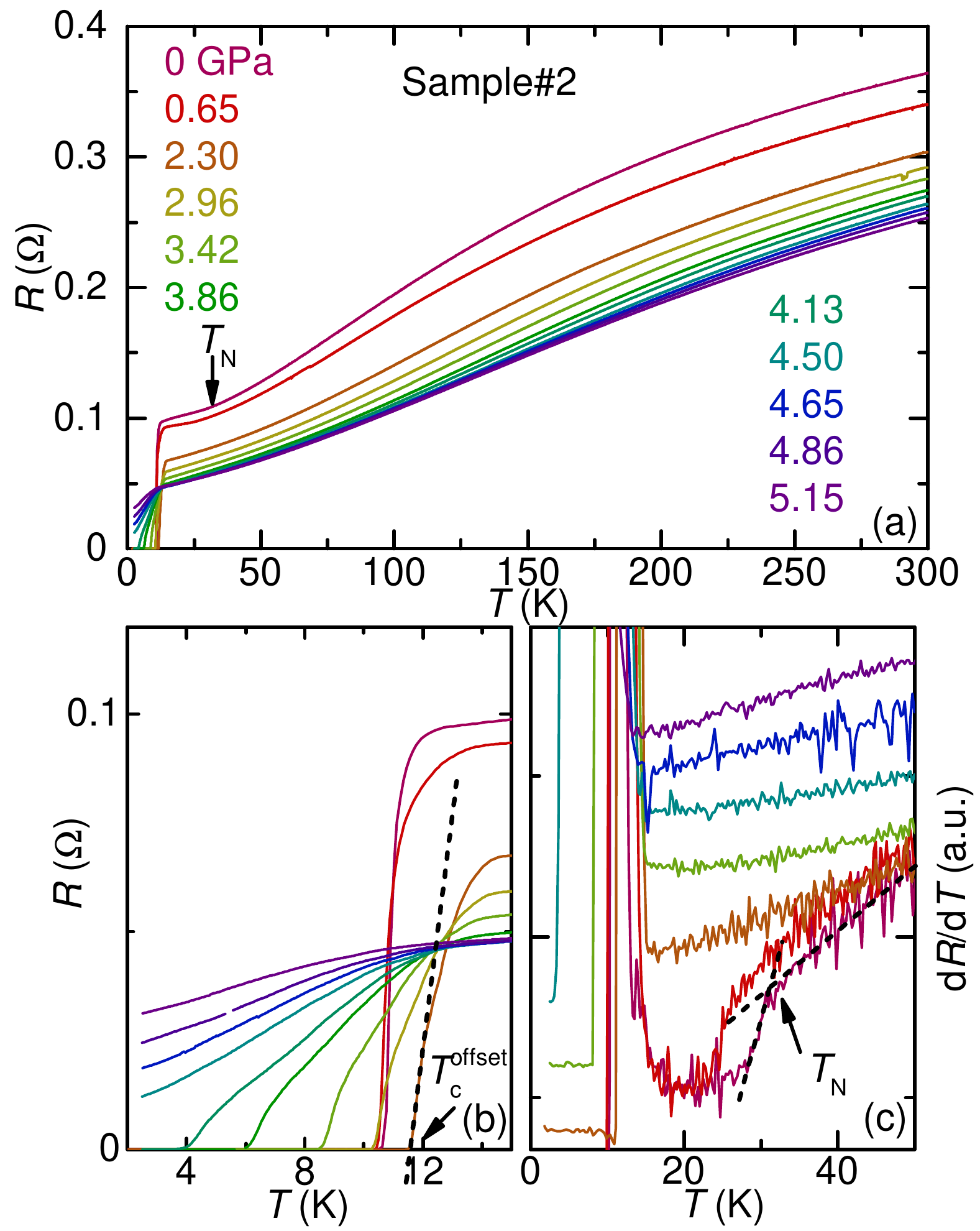} 
			\caption{(a) Evolution of the in-plane resistance with hydrostatic pressure up to 5.15 GPa measured in a modified Bridgman anvil cell (MBAC) for CaK(Fe$_{0.976}$Mn$_{0.024}$)$_4$As$_4$, sample 2. (b) Enlarged view of the low-temperature resistance showing the superconducting transition. Criterion for $T_c^{offset}$ is indicated by dashed line and arrow in the figure. (c) Temperature derivative of the resistance, d$R$/d$T$, showing the evolution of the magnetic transition $T_{N}$. Criterion for $T_{N}$ is indicated by dashed lines and arrow in the figure.}
			\label{fig2}
		\end{center}
	\end{figure}

	Figures \ref{fig1}(a) and \ref{fig2}(a) present the temperature dependent resistance of CaK(Fe$_{0.976}$Mn$_{0.024}$)$_4$As$_4$ sample 1 and 2 in a PCC under pressures up to 2.03 GPa and in a MBAC under pressures up to 5.15 GPa, respectively (Figs. \ref{fig1}(c) and \ref{fig2}(c)). For both samples, at ambient pressure, resistance decreases upon cooling, showing metallic behavior. At $\sim$ 30 K, a kink-like anomaly is observed and is associated with a magnetic transition $T_ N$. At $\sim$ 10 K, a sharp drop of resistance to zero is observed (Figs. \ref{fig1}(b) and \ref{fig2}(b)) as the compound goes through the superconducting transition\cite{Xu2022PRB}. 
	
	Upon increasing pressure several changes take place. The normal state resistance decreases and $T_ N$ and $T_ c$ change as shown in Figs. \ref{fig1} and \ref{fig2}. As shown in the figures, $T_ c$ increases upon increasing pressure up to $\sim$ 2 GPa. Further increasing of pressure suppresses $T_ c$. In addition, the sharp superconducting transition at low pressures becomes broadened at higher pressures. For $p\ge$ 4.50 GPa, the resistance does not drop to zero down the lowest temperature measured in this study (1.8 K).
	
	To better trace the magnetic transition temperature $T_ N$, the temperature derivative, d$R$/d$T$ is calculated and plotted in Figs. \ref{fig1}(c) and \ref{fig2}(c). It is seen that the kink-like anomaly in $R(T)$ is revealed as a step-like anomaly in d$R$/d$T$. As shown in the figures, $T_ N$ is suppressed by increasing pressure until it reaches the $T_ c$($p$) line and is not resolved any more at higher pressures.
	
\begin{figure}
		\begin{center}
			\includegraphics[width=\columnwidth]{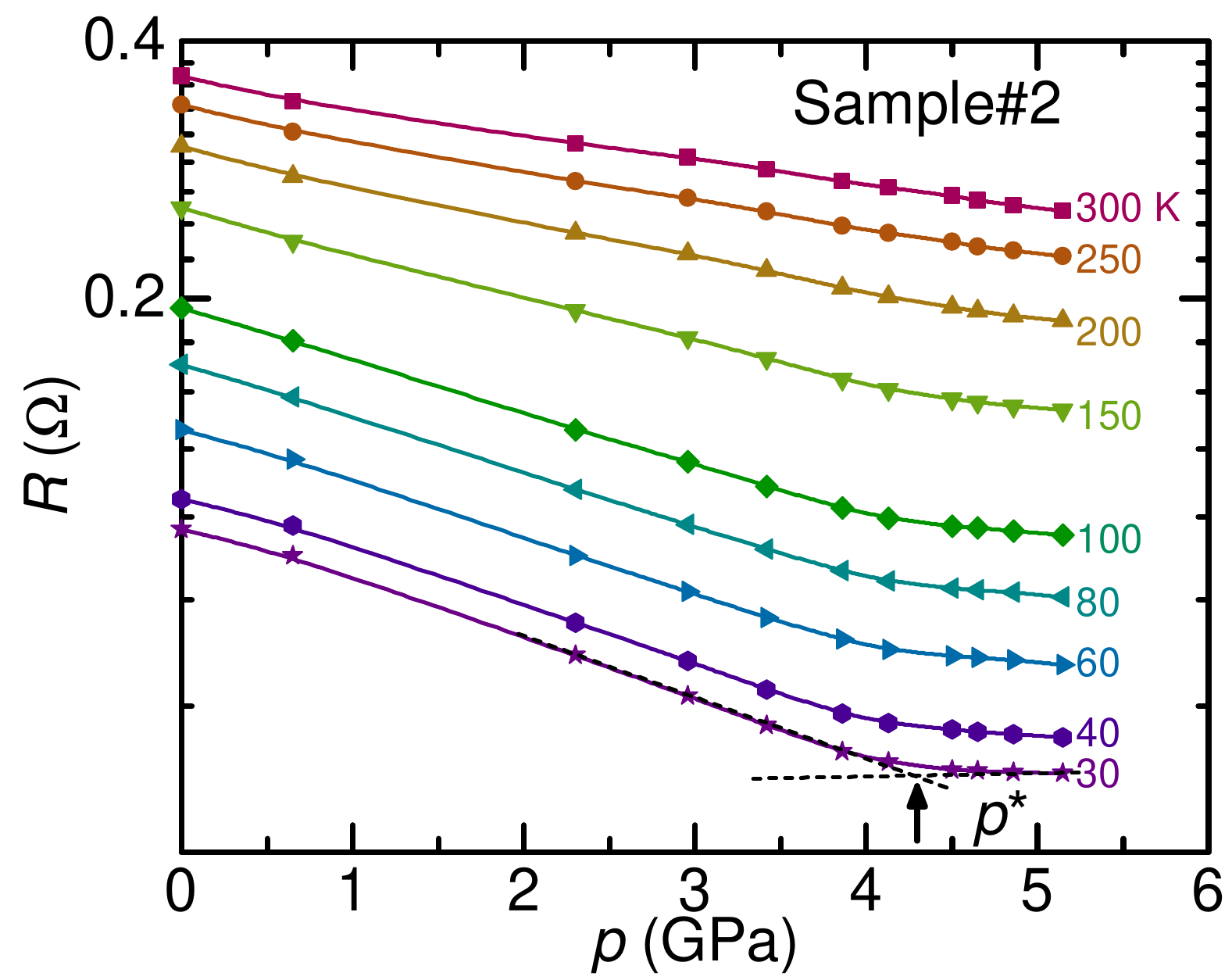} 
			\caption{Pressure dependence of the resistance, $R(p)$, at fixed temperatures for CaK(Fe$_{0.976}$Mn$_{0.024}$)$_4$As$_4$ sample 2. The critical pressure $p^*$ (indicated by dashed lines and arrow), which is associated with the hcT phase, is described in details in the text.}
			\label{fig3}
		\end{center}
	\end{figure}

	The pressure dependent resistance at fixed temperatures for sample 2 is further analyzed and plotted in Fig. \ref{fig3}. Again, at all temperatures above superconducting transition, resistance decreases upon increasing pressure. In addition, a kink-like anomaly in the $R(p)$ curves is observed at a critical pressure $p^*\sim$ 4.3 GPa. It is worth noting such a behavior is also observed in the parent and Ni-substituted CaKFe$_4$As$_4$\cite{Kaluarachchi2017PRB,Xiang2018PRB,Gati2020a}, where the anomaly is associated with a half-collapsed-tetragonal (hcT) phase transition at around 4-4.15 GPa. In the parent CaKFe$_4$As$_4$ compound, it is demonstrated that across the hcT transition, the As-As bonds forms across the Ca layer, which causes a sudden decrease in the $c$ lattice parameter and an expansion in the $a$ lattice parameter\cite{Kaluarachchi2017PRB}. In addition such a collapsed tetragonal transition is often accompanied by significant changes of electronic properties which can lead to the loss of superconductivity\cite{Kaluarachchi2017PRB,Borisov2018,Gati2012PRB}. Based on the analogy, we identify this anomaly as an indication of the hcT transition that exists from base temperature up to at least 30 K. The superconducting transition for $p>p^*$ is likely not bulk, which could cause the broadening of the superconducting transition at high pressures as seen in Fig. \ref{fig2}(a)\cite{Kaluarachchi2017PRB,Xiang2018PRB}. Similar hcT transitions are also predicted and observed in other $AeA$Fe$_4$As$_4$ compounds\cite{Borisov2018}. Specifically, the hcT transition for CaRbFe$_4$As$_4$, RbEuFe$_4$As$_4$ and CsEuFe$_4$As$_4$ are experimentally shown to be $\sim$ 6GPa, $\sim$ 10 GPa and $\sim$ 12 GPa\cite{Stillwell2019,Jackson2018}. Note that in the parent and Ni-substituted CaKFe$_4$As$_4$, the hcT transition remains around 4 GPa, demonstrating the insensitivity of the hcT transition pressure to Ni or Mn substitution.
	
\begin{figure}
		\begin{center}
			\includegraphics[width=\columnwidth]{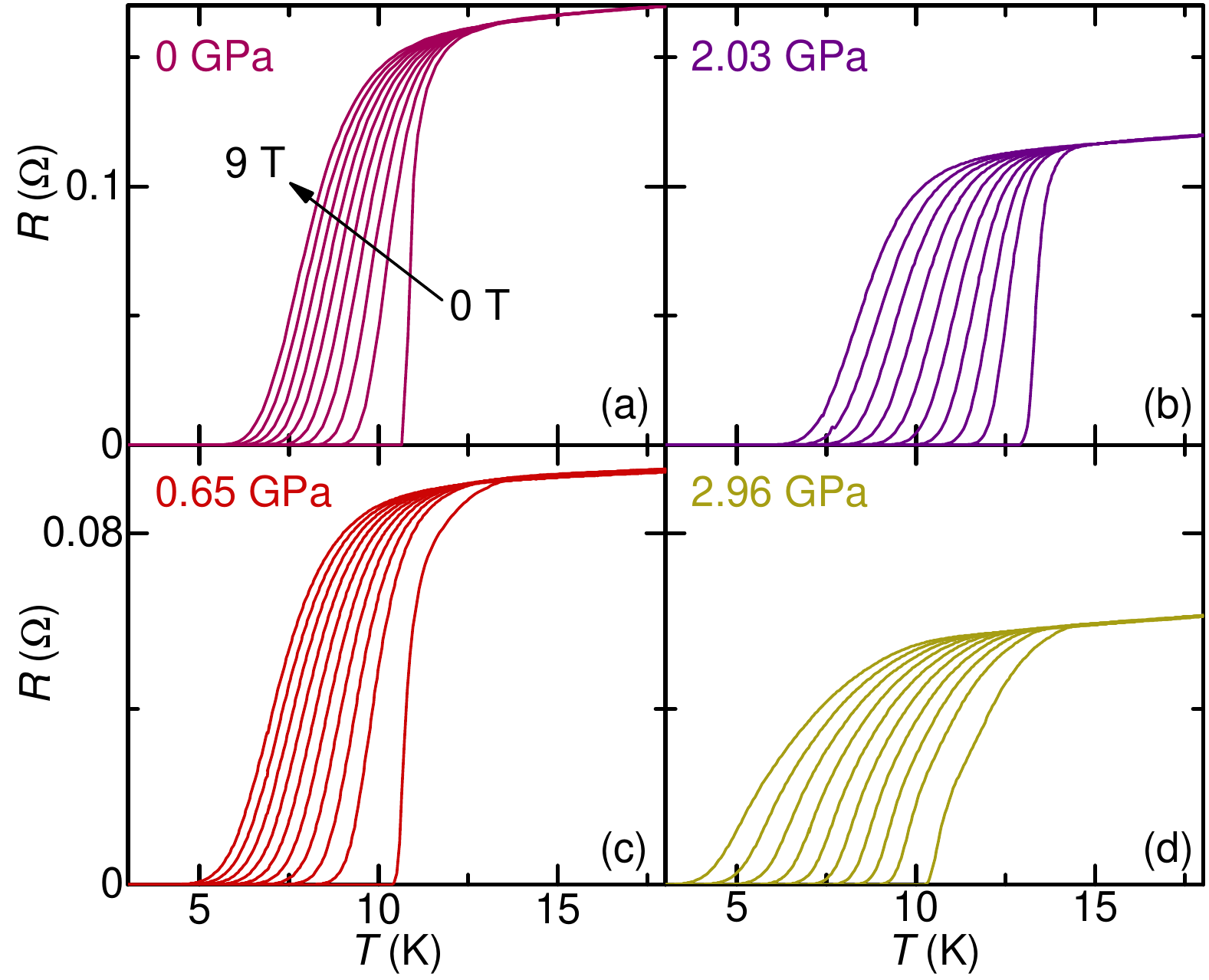} 
			\caption{Temperature dependence of resistance under magnetic field up to 9 T for selected pressures for CaK(Fe$_{0.976}$Mn$_{0.024}$)$_4$As$_4$ sample 1 ((a) and (b)) and sample 2 ((c) and (d)).}
			\label{fig4}
		\end{center}
	\end{figure}

\begin{figure}
		\begin{center}
			\includegraphics[width=\columnwidth]{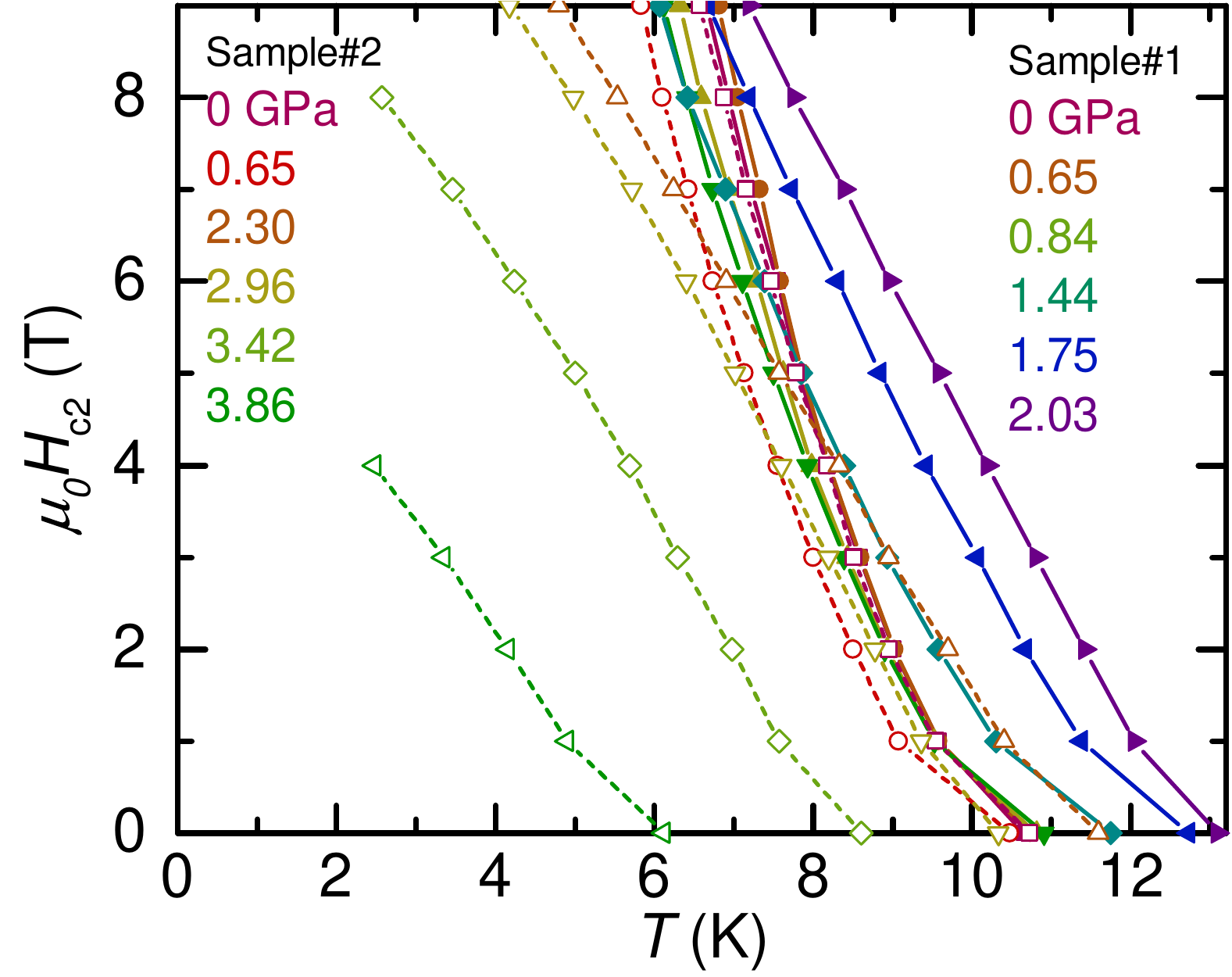} 
			\caption{Temperature dependence of the superconducting upper critical field $H_{c2}(T)$ under selected pressures for CaK(Fe$_{0.976}$Mn$_{0.024}$)$_4$As$_4$. $T_c^{offset}$ is used. Solid and open symbols are two samples measured in PCC and MBAC, respectively.}
			\label{fig5}
		\end{center}
	\end{figure}

	The superconducting upper critical field $H_{c2}$ can be evaluated from the $R(T)$ data taken at different applied magnetic fields and pressures, shown in Fig. \ref{fig4} at pressures lower than $p^*$, where the superconductivity is believed to be bulk, using the offset criteria defined in Figs. \ref{fig1} and \ref{fig2}. The temperature dependent $H_{c2}$ under pressures up to 3.86 GPa is presented in Fig. \ref{fig5}. As shown in the figure, $H_{c2}$ is linear in temperature except for magnetic fields below 1 T. This curvature at low magnetic fields has been also observed in other FeSC and can be explained by the multiband nature of superconductivity\cite{Kogan2012,Kaluarachchi2016,Xiang2017PRB,Mou2016PRL}. In addition, it is observed that the evolution of the temperature-dependent $H_{c2}$ with pressure is nonmonotonic.
	
\begin{figure}
		\begin{center}
			\includegraphics[width=\columnwidth]{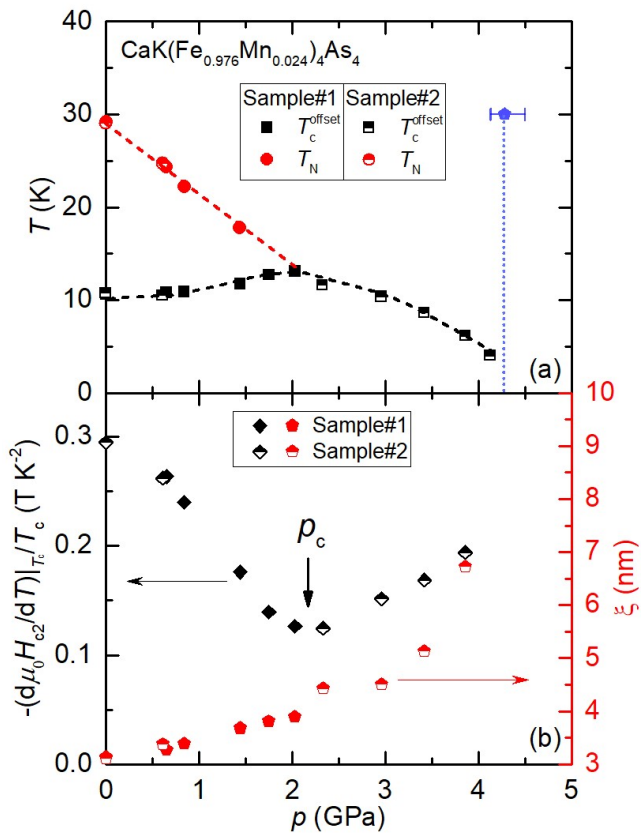} 
			\caption{(a) Pressure-temperature phase diagram of CaK(Fe$_{0.976}$Mn$_{0.024}$)$_4$As$_4$ as determined from resistance measurements. The squares and circles represent the superconducting $T_c^{offset}$ and magnetic $T_ N$ phase transitions.The blue dotted line indicates the half-collapsed-tetragonal phase transition up to 30 K, inferred from the pressure-dependent resistance $R(p)$ data in Fig. \ref{fig3}. (b) Pressure dependence of the normalized upper critical field slope -(1/$T_ c$)(d$\mu_0 H_{c2}$/d$T$)$\vert_{T_ c}$ (black symbols) and inferred coherence length $\xi$ (red symbols). A local minimum in the slope and a clear change of the coherence length at $p_c$ (indicated by the arrow) are observed near the pressure where $T_c^{offset}$$(p)$ and $T_ N$$(p)$ lines intersect.}
			\label{fig6}
		\end{center}
	\end{figure}

	The evolution of the superconducting and the magnetic transition temperatures with pressure is summarized in a pressure-temperature phase diagram presented in Fig. \ref{fig6}. $T_c^{offset}$ and $T_ N$ values are obtained using the criteria defined in Figs. \ref{fig1} and \ref{fig2} and the hcT phase above $p^*$ is indicated by the blue dotted line in the figure. We see that for CaK(Fe$_{0.976}$Mn$_{0.024}$)$_4$As$_4$, upon increasing pressure from 0 to $\sim$ 1.5 GPa, $T_ N$ is monotonically suppressed from 30 K to 18 K. In terms of the superconducting transition temperatures, $T_c^{offset}$ first increases from 10 K to 13 K with pressure increasing from 0 to 2.03 GPa. Upon further increasing pressure to 4.13 GPa, $T_c^{offset}$ is suppressed to 3.9 K, resulting in a local maximum in $T_ c$ at $\sim$ 2 GPa. Above $p^*\sim$ 4.3 GPa, CaK(Fe$_{0.976}$Mn$_{0.024}$)$_4$As$_4$ enters into the hcT phase at low temperatures. The pressure-temperature phase diagram of CaK(Fe$_{0.976}$Mn$_{0.024}$)$_4$As$_4$ is qualitatively similar to that of CaK(Fe$_{1-x}$Ni$_{x}$)$_4$As$_4$ ($x$ = 0.033, 0.05)\cite{Xiang2018PRB}. For all three compounds, The $T_ N$($p$) and $T_ c$($p$) lines intersect at the maximum $T_ c$ point of the $T_ c$($p$) curve. Note that below we call this intersect-pressure $p_ c$ (identified by a sharp minima in the  -(1/$T_ c$)(d$\mu_0 H_{c2}$/d$T$)$\vert_{T_ c}$ data)\cite{Xiang2018PRB}. Quantitatively, the suppression of $T_ N$ with $p$ is faster for Mn-substitution, compared with that for Ni-substitution (the initial suppression of $T_ N$ is $\sim$ 8 K/GPa for Mn-substitution, $\sim$ 3.3 K/GPa and $\sim$ 6.5 K/GPa for Ni-substitution $x$ = 0.033 and 0.05, correspondingly). In addition, $p_ c$ for Mn-substitution ($\sim$ 2 GPa) is lower than that of Ni-substitution ($\sim$ 3 GPa for $x$=0.033 and 0.05)\cite{Xiang2018PRB}. Such non-monotonic behavior of $T_ c$ under pressure could be qualitatively understood as a result of coexistence of  itinerant magnetism and superconductivity that are competing in the same, shared, electron subsystem\cite{Machida1981,Budko2018,Kreyssig2018}.
	
	The superconducting upper critical field $H_{c2}$ was further analyzed following Refs. \onlinecite{Xiang2018PRB,Kaluarachchi2016,Xiang2017PRB,Taufour2014}. Generally speaking, the slope of the upper critical field normalized by $T_ c$ is related to the Fermi velocity and superconducting gap of the system\cite{Kogan2012}. In the clean limit for a single band case,
	\begin{equation}\label{eq:1}
		-(1/T_{ c})(d\mu_0 H_{{c2}}/dT)\vert_{T_{ c}}\propto 1/v_{F}^2
	\end{equation}
	where $v_{F}$ is the Fermi velocity. Even though the superconductivity in CaKFe$_4$As$_4$ is multiband, Eq.\ref{eq:1} can give qualitative insight into changes induced by pressure. As shown in Fig. \ref{fig6}(b), upon increasing pressure, the normalized slope of the upper critical field -(1/$T_ c$)(d$\mu_0 H_{c2}$/d$T$)$\vert_{T_ c}$ (the slope d$\mu_0 H_{c2}$/d$T\vert_{T_ c}$ is calculated by linear fitting the $H_{{c2}}(T)$ data above 1 T in Fig. \ref{fig5}) first decreases and then increases, resulting a minimum of -(1/$T_ c$)(d$\mu_0 H_{c2}$/d$T$)$\vert_{T_ c}$ at $p_c$ in the studied pressure range. In addition, $p_c$ coincide with the cross of the $T_{N}$($p$) and $T_{c}$($p$) lines. Note that similar upper-critical-field behavior is also observed in the Ni-substituted CaK(Fe$_{1-x}$Ni$_x$)$_4$As$_4$\cite{Xiang2018PRB}. 
	
	Coherence length, $\xi$, can be estimated from $|$d$ H^{\parallel c}_{c2}$/d$T|$=$\phi_0/2\pi\xi^2_\perp T_ c$, where $\phi_0$ is the magnetic flux quantum, $\xi_\perp$ is the effective Ginzburg-Landau coherence length near $T_ c$\cite{Gurevich2011,Xu2022PRB} when magnetic field is applied perpendicular to the $ab$ plane. As shown in the figure, $\xi$ monotonically increases with increasing pressure. In addition, $\xi$($p$) shows a sudden increase at $p_c$. This is not surprising as coherence length $\xi$ is also inferred from the upper critical field slope, d$H_{c2}$/d$T$.
	
	The clear changes in the pressure dependence of normalized slope  upper critical field slope, -(1/$T_ c$)(d$\mu_0 H_{c2}$/d$T$)$\vert_{T_ c}$, and in coherence length, $\xi$, coincides with the pressure, $p_c$, across which $T_{N}$ is suppressed below $T_{c}$. We argue that this indicates a possible Fermi-surface reconstruction, which could happen due to the pressure-induced disappearance of magnetism, which is also observed in the BaFe$_2$As$_2$ family\cite{Liu2009PRB,Dai2009,Liu2010a,Liu2011,Xiang2018PRB}.

	\section{Conclusion}
	In summary, the resistance of the Mn-substituted iron-based superconductor CaK(Fe$_{1-x}$Mn$_x$)$_4$As$_4$ (x = 0.024) was investigated under hydrostatic pressures up to 5.15 GPa and in magnetic fields up to 9 T. It is observed that the magnetic transition temperature $T_ N$ is suppressed upon increasing pressure, whereas the superconducting transition temperature exhibits a nonmonotonic dependence on pressure with a maximum at $\sim$ 2 GPa in the studied pressure range. CaK(Fe$_{1-x}$Mn$_x$)$_4$As$_4$ (x = 0.024) likely goes through a half-collapsed-tetragonal phase transition when pressure is increased across $\sim$ 4.3 GPa. A minimum in the normalized  slope of the upper critical field, -(1/$T_ c$)(d$\mu_0 H_{c2}$/d$T$)$\vert_{T_ c}$, and a sudden change of the coherence length, $\xi$, are observed at a pressure where $T_{N}$($p$) and $T_{c}$($p$) lines intersect. This suggests a possible Fermi-surface reconstruction associated with the disappearance of the magnetism.
	
	The results of these studies, show that there is no obvious or qualitative difference in the pressure response between Ni- and Mn-substituted CaKFe$_4$As$_4$. The observation of lower value of $p_c$ in Mn-substituted CaKFe$_4$As$_4$ is, at least in part, due to lower $T_N$ at ambient pressure. We do not see unambiguous, additional, contribution from AG pair-breaking to the evolution of $T_c$ under pressure, in contrast to the results on borocarbides\cite{Michor2000,Choi2001}, but in agreement with the behavior of Gd-substituted La$_3$In\cite{Maple1969}. As such then, the pressure dependent phase diagram for the Mn-substituted CaKFe$_4$As$_4$ provides a full set of benchmarks that any theory trying to understand these materials would need to accommodate.
	
	\section{Acknowledgements}
	This work is supported by the U.S. DOE, Basic Energy Sciences, Materials Science and Engineering Division under Contract No. DE-AC02-07CH11358. L. X. was supported by the Gordon and Betty Moore Foundation’s EPiQS Initiative through Grant No. GBMF4411 and, in part, by the W. M. Keck Foundation.

	\clearpage
	\bibliographystyle{apsrev}

\end{document}